\documentclass[sigconf,screen,nonacm]{acmart}
\setcopyright{none}
\AtBeginDocument{%
  }

\setcopyright{acmlicensed}
\copyrightyear{2026}
\acmYear{2026}
\acmDOI{XXXXXXX.XXXXXXX}
\acmConference[CHASE '26]{The 19th International Conference on Cooperative and Human Aspects of Software Engineering}{April 13--14,
  2026}{Rio de Janeiro, Brazil}
\acmISBN{978-1-4503-XXXX-X/2018/06}

\usepackage{mdframed}
\usepackage{soul}

\usepackage{graphicx}
\usepackage{booktabs}
\usepackage{framed}
\usepackage{changepage}
\usepackage{tcolorbox}

\usepackage{changepage}
\newcommand{\interviewquote}[1]{
 \def\FrameCommand{
    \hspace{0pt} 
    {\color{cyan}  \vrule width 2pt}%
    {\color{white} \vrule width 3pt}%
  }%
  \MakeFramed{\advance\hsize-\width\FrameRestore}%
  \noindent
  \begin{adjustwidth}{}{1pt}
  {\footnotesize``#1''}
  \end{adjustwidth}
  \endMakeFramed%
}

\begin{document}

\title{Folklore in Software Engineering: \\ A Definition and Conceptual Foundations}

\author{Eduard Paul Enoiu}
\email{eduard.paul.enoiu@mdu.se}
\orcid{0000−0003−2416−4205}
\author{Jean Malm}
\email{jean.malm@mdu.se}
\orcid{1234-5678-9012}
\affiliation{%
  \institution{M\"{a}lardalen University}
  \city{V\"{a}ster\r{a}s}
  \country{Sweden}
}

\author{Gregory Gay}
\email{greg@greggay.com}
\orcid{0000-0001-6794-9585}
\affiliation{%
  \institution{Chalmers University of Technology and \\ University of Gothenburg}
  \city{Gothenburg}
  \country{Sweden}
}

\renewcommand{\shortauthors}{E. P. Enoiu et al.}

\begin{abstract}
We explore the concept of folklore within software engineering, drawing from folklore studies to define and characterize narratives, myths, rituals, humor, and informal knowledge that circulate within software development communities.
Using a literature review and thematic analysis, we curated exemplar folklore items (e.g., beliefs about where defects occur, the $10x$ developer legend, and technical debt). We analyzed their narrative form, symbolic meaning, occupational relevance, and links to knowledge areas in software engineering. 
To ground these concepts in practice, we conducted semi-structured interviews with 12 industrial practitioners in Sweden to explore how such narratives are recognized or transmitted within their daily work and how they affect it. Synthesizing these results, we propose a working definition of software engineering folklore as informally transmitted, traditional, and emergent narratives and heuristics enacted within occupational folk groups that shape identity, values, and collective knowledge. 
We argue that making the concept of software engineering folklore explicit provides a foundation for subsequent ethnography and folklore studies and for reflective practice that can preserve context-effective heuristics while challenging unhelpful folklore. 
\end{abstract}

\maketitle

\section{Introduction}

\begin{figure*}[!t]
\centering
\includegraphics[width=0.77\textwidth]{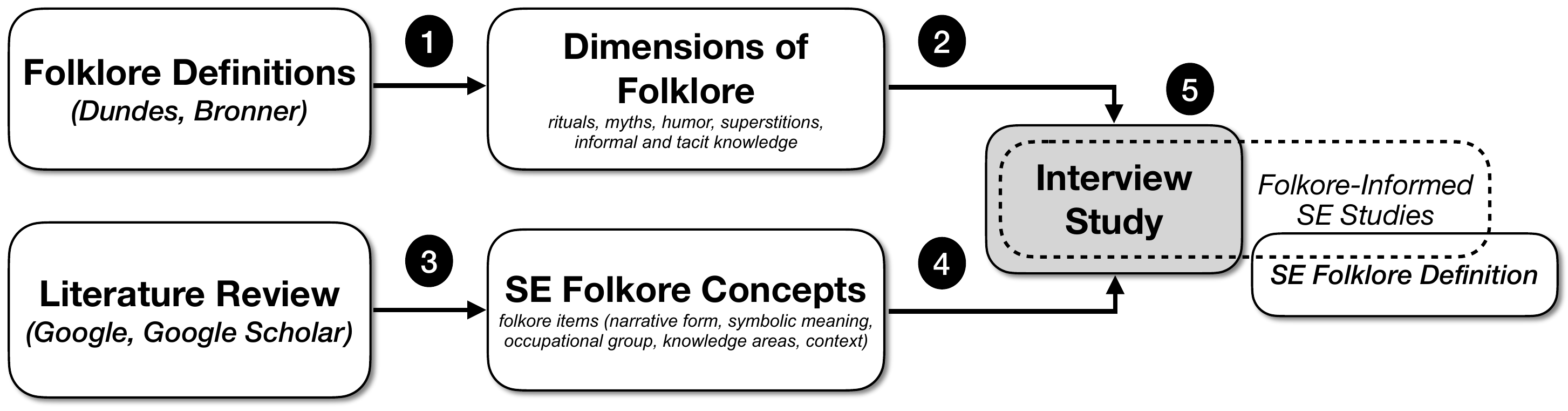}
\Description{Visualization of the method used for exploring SE folklore, described in the text.}
\caption{Overview of the method used to explore SE folklore: theory grounding, folklore items, practitioner interviews, and synthesis into a working definition.}
\label{fig:se-folklore-method} 
\end{figure*}

Software Engineering (SE) has generated a body of shared narratives, beliefs, jokes, customs, and myths, much like other established fields or communities~\cite{shull2012ibelieve}. From stories to narratives, these cultural artifacts seem to be passed among developers, testers, and managers. These narratives range from historical anecdotes and rituals to humor and persistent myths about productivity and development practices. Even if we often dismiss them as anecdotal or unscientific, such shared narratives appear to play a significant role in shaping how software professionals perceive their work, make decisions, and construct identities~\cite{bossavit2012leprechauns,mendez2021interdiscipline,zagalsky2018knowledge}.

Folklore, in the academic sense, refers to the collective traditions, stories, and customs of a group of people. Folklorists, such as Alan Dundes~\cite{4743c5ea-ef11-30f3-a9c2-5db3158de407,dundes2019folk} and Simon J. Bronner~\cite{bronner2016folklore}, emphasize that folklore is alive and evolving, even in modern settings. Dundes defined ``folk'' as any group of people who share at least one common factor---be it an occupation, language, or any shared identity---and noted that any such group will have traditions it calls its own. 

Every culture has origin stories and legends---software development is no different. Consider the ``first computer bug'', recorded in 1947 \cite{beyer2012grace,kidwell2002stalking}, where engineers found a moth stuck in a Harvard Mark II relay and taped it in the logbook with the note ``First actual case of bug being found''. This story, often attributed to computing pioneer Grace Hopper\footnote{See Smithsonian National Museum of American History, Log book with computer bug, \url{https://n2t.net/ark:/65665/ng49ca746a3-b8b7-704b-e053-15f76fa0b4fa}}, has been retold for decades as the origin of the term ``debugging''. The artifact (the moth in the logbook) is preserved by a museum and is an iconic piece of technology lore. 

Just as folklore includes legends, it also contains myths and beliefs---things widely repeated that may be oversimplified, exaggerated, or outright false. SE has plenty of these ``received wisdoms circulating among teams''. For example, a persistent belief is that some developers are 10 times more productive than others~\cite{bossavit2012leprechauns}. This notion stems from claims like ``the best programmers are up to 28 times better than the worst programmers''. Such claims, cited as ``accepted truths'' by authors like Robert Glass~\cite{glass2002facts} and Laurent Bossavit~\cite{bossavit2012leprechauns}, have gained almost legendary status.  



In SE literature (e.g.,~\cite{bossavit2012leprechauns,mendez2021interdiscipline,glass2002facts}), there is a tradition of identifying and debunking software myths, essentially exposing beliefs that does not hold up to scrutiny. Laurent Bossavit's ``The Leprechauns of Software Engineering''~\cite{bossavit2012leprechauns} explicitly calls out how folklore can turn into facts in the field through constant repetition and variation. Prior research has also examined how practitioners share experience through narratives such as war stories \cite{lutters2007revealing,reifer2013software} and how everyday work departs from formal processes \cite{boden2012knowledge,ingram2020software,meyer2019today}, as also emphasized in ethnographic investigations (e.g., plans and situated actions \cite{suchman1987plans}). Other lines of work in knowledge engineering and knowledge management have also focused on eliciting and externalizing expertise (e.g., \cite{ryan2013acquiring}).

This study seeks to explore the concept of SE folklore by drawing from traditional folklore theory and applying it to the occupational practices within software development:
\newline
\begin{mdframed}[linewidth=1pt, linecolor=black, backgroundcolor=gray!10, roundcorner=5pt]
\textbf{What constitutes SE folklore, and how can it be defined?}
\end{mdframed}
\vspace{2mm}
We aim to define what constitutes folklore in SE, identify its recurring dimensions, and examine how it manifests in different roles and knowledge areas. We examine existing implicit and explicit characterizations. We also seek to identify and categorize specific manifestations of folklore within SE practices. Ultimately, we define the concept of SE folklore and propose several areas of investigation within the context of SE. 
By synthesizing concepts from folklore studies with SE research, we cast a spotlight on a largely overlooked aspect of the field, and lay the foundation for future investigations into how cultural narratives shape SE.

\section{Research Method}

We aim to define and characterize SE folklore by identifying its dimensions and manifestations in SE contexts. The research process followed is illustrated in Figure~\ref{fig:se-folklore-method}.

\subsection{Folklore Definitions and Dimensions}
A comparative synthesis of folkloristics and folklore studies over time (e.g., \cite{ben1971toward,hartland1891report,jacobs1893folk,thompson1951folklore}) highlights different perspectives, including traditional knowledge, performance and event analysis, communication and genre approaches, and material culture traditions. With this intellectual legacy in mind, we focus on the modern reformulation of folklore \cite{bronner2016folklore,bronner2016toward} that emphasizes social interaction, a process-oriented conception of tradition, living folklore, and the roles of cognition and practice, as these lenses are the most immediately applicable to occupational folklore in SE settings.

We apply the definitions and categories from folklore studies---particularly from Dundes~\cite{4743c5ea-ef11-30f3-a9c2-5db3158de407} and  Bronner~\cite{bronner2016folklore}---to explore folklore within an SE context (Step 1 in Figure~\ref{fig:se-folklore-method}). Dundes frames folklore as informally transmitted traditional knowledge shared among members of a folk group, emphasizing the roles of communication, repetition, and group identity. Bronner expands this by viewing folklore as embodied in everyday routines, rituals, artifacts, and symbols, underscoring its cultural and practical dimensions. Building on these foundations, we adapted dimensions such as rituals, myths, humor, superstitions, and informally transmitted knowledge as potential indicators of folklore (Step 2). These categories guide the development of a working definition of SE folklore.

\subsection{Software Engineering Folklore Concepts}\label{sec:lit}

To explore how these folkloric dimensions manifest in practice (Step 3 in Figure~\ref{fig:se-folklore-method}), we manually collected and analyzed articles that either explicitly or implicitly engage with beliefs, informal knowledge, rituals, or cultural practices in software development. 

One author conducted the search and initial screening; the resulting set of publications and items was then iteratively discussed with the other authors to refine the analysis. The initial classification was performed by the same author and reviewed and aligned through discussion with the author team. The selection process was informal and exploratory. We used Google Scholar and the Google search engine to identify literature that mentioned folklore in the context of SE or addressed relevant themes, using the phrase ``\textit{software engineering folklore}'' and screening results by title. This process resulted in a set of items that should be interpreted as illustrative rather than comprehensive. 

We then identified an example set of SE folklore concepts that characterize specific folklore items (Step 4 in Figure~\ref{fig:se-folklore-method}). To analyze these items, we used interpretive concepts (i.e., form, performance, meaning, transmission) informed by Bronner’s practice-oriented description of folklore as cognitive and communicative praxis~\cite{bronner2016toward}, as well as the \textit{narrative forms} (e.g., myth, legend, anecdote, humor, ritual, artifact) drawn from Alan Dundes’ typology~\cite{4743c5ea-ef11-30f3-a9c2-5db3158de407}: \newline
\begin{mdframed}[linewidth=1pt, linecolor=black, backgroundcolor=gray!10, roundcorner=5pt]
\textbf{Myths} are stories believed to be true by a group.

\smallskip\noindent \textbf{Legends} are stories with a historical basis.

\smallskip\noindent \textbf{Anecdotes} are illustrative stories based on experience.

\smallskip\noindent \textbf{Rituals} are symbolic and repeated performed actions.

\smallskip\noindent \textbf{Artifacts} are material or symbolic objects.
\end{mdframed}
\vspace{5pt}

We classify the narrative form of each item, as well as its \textit{symbolic meaning}, related \textit{occupational groups}, and relevant \textit{knowledge areas}.
    The ``\textit{symbolic meaning}'' refers to the cultural depth of that item, as outlined in Schein's model of organizational culture~\cite{schein2010organizational}. This concept captures whether the folklore functions as an artifact, an espoused value, or a basic underlying assumption\footnote{Folklore items were interpreted as artifacts (visible behaviors, structures, processes), espoused beliefs and values (stated ideals, goals, justifications), or basic underlying assumptions (taken for granted beliefs shaping perception and action).}.
    Each symbolic layer represents a different degree of cultural visibility and resistance to change, and each interacts differently with folklore.
    The \textit{occupational group} identifies the professional roles most closely associated with the folklore item, such as developers, testers, maintainers, architects, or project managers.
    Finally, each folklore item is mapped to one or more \textit{SE knowledge areas} using the SWEBOK classification \footnote{For more details, we refer the reader to the SWEBOK guide: \url{https://sebokwiki.org/wiki/An_Overview_of_the_SWEBOK_Guide}}, thereby linking cultural expressions to specific technical domains, such as testing, maintenance, or design. 

\subsection{Interview Study}

Drawing on SE concepts and dimensions of folklore, we designed an interview study to explore practitioner experiences (Step 5 in Figure~\ref{fig:se-folklore-method}). Through \textit{iterative analysis and thematic coding}, we expanded the concept of SE folklore in practice and developed a working definition for future studies.

\subsubsection{\textbf{Data Collection}} 
We conducted face-to-face and online \textit{semi-structured interviews}, following the guidelines by Lin\r{a}ker et al.~\cite{linaaker2015guidelines}. We follow Strandberg's guidance on ethical interviews~\cite{strandberg2019ethical}: consent was obtained, raw audio was stored with time and space limits, and transcripts were anonymized. 

The \textit{interview protocol} was planned and refined through piloting (in three initial interviews). Our analytic stance was semantic, emphasizing broad and data-driven coding while also attempting to fit the responses into the prior theoretical frame presented in Section~\ref{sec:lit}. Before each interview, we explained the purpose, emphasized voluntary participation and the right to skip any question or withdraw, and described confidentiality practices. We requested permission to audio record solely to ensure accuracy; when granted, recordings were kept confidential. Interviews began with a brief icebreaker where participants described their current role, tenure in SE, and organizational background.

The \textit{interview guide} included three main sections (see Table \ref{tab:interview_questions}). Section 1 produced \textit{general perceptions} by asking participants to define folklore in SE in their own context. Optional prompts included informally transmitted knowledge (stories, sayings, jokes, traditions) and recurring practices or rituals. Section 2 investigated four folklore types---myths and beliefs (including rules of thumb around productivity, quality, testing, programming, and management, and whether such beliefs are questioned or taken for granted); anecdotes and legends (repeated stories about SE projects, engineers, or memorable bugs, and the roles such stories play); rituals and practices (recurring SE routines such as stand-ups or events and any meanings beyond their practical purposes, e.g., identity or value signaling); and artifacts and humor (objects, memes, jokes, and what they imply about group identity, culture, or values in SE). Finally, Section 3 focused on transmission and impact, exploring how newcomers adopt folklore, how it endures over time, whether it has influenced decisions or conflicts, how its expression varies across organizations, and whether its effects are perceived as helpful, harmful, or mixed. Each interview closed with an invitation for additional examples and a debrief thanking the participant.

A pair of researchers conducted sessions---a primary interviewer and a second researcher who took notes and posed follow-up questions. Interviews typically lasted 45 to 60 minutes. Where recording was possible, we produced verbatim transcripts. When recording was not feasible, we created detailed summary transcripts that preserved the substance and meaning of the exchange and took contemporaneous or immediate post-hoc notes. All material was anonymized during transcription; personally identifying details and organizational names were replaced with neutral placeholders. Participants reviewed their transcripts for accuracy before analysis.

\begin{table}[!t]
    \centering
    \scriptsize
    \caption{Interview guide.}\label{tab:interview_questions} 
    \begin{tabular}{p{0.45\textwidth}}
    \toprule
   1. Could you briefly describe your current role and how long you have been working in SE? \\
2. How many different software dev organizations have you worked for? \\
    \midrule
   3. In a general sense, how would you define the term folklore in software engineering?\\ 
4. When you think about folklore in the context of your work, what comes to mind?\\ 
    \midrule
5. Have you heard any myths and persistent beliefs, or any rules of thumb (e.g., about productivity, quality, testing, programming, management)? \\
6. Are there stories in your team, company, or community that people tend to repeat, for example, about past projects, engineers, or memorable bugs?\\
7. Can you describe any recurring practices or rituals you have encountered (for example, stand-ups, recurring events, or other routines)? \\
8. Are there objects, memes, or jokes that circulate within your team, company, or community?\\
    \midrule
  9. Can you recall how this folklore was passed on to you when you joined a team or company? \\
10. Have you ever seen or experienced a piece of folklore influence a decision, shape an attitude, or contribute to a conflict at work? \\
11. How does the way folklore (such as beliefs, stories, rituals, or humor) is expressed or treated differ between the organizations you have worked in? \\
12. In your experience, do you see folklore as mostly helpful, mostly harmful, or a mix of both? \\
 \midrule
13. Before we wrap up, is there anything else you would like to share, any examples or experiences we have not touched on, that would be important for this study?  \\
    \bottomrule
    \end{tabular} 
\end{table}

\subsubsection{\textbf{Population and Sampling}} 

We used \textit{purposive convenience sampling} to obtain variation in roles and organizational contexts among practitioners based in Sweden. For this exploratory phase, we aimed for 10--15 interviews to achieve depth and role diversity, monitor for thematic saturation, and to be able to add cases incrementally if new themes continued to emerge. We recruited practitioners through professional and research networks via direct invitations, and iteratively searched additional participants to address gaps in gender, role, and organizational context; recruitment stopped at $n=12$ when successive interviews reiterated similar topics and no substantially new themes emerged.

\begin{table}[!t]
    \centering
    \scriptsize
    \caption{Participant IDs and demographics.}\label{tab:participants_demo}
    \begin{tabular}{p{0.02\textwidth} p{0.40\textwidth}}
    \toprule
    \textbf{ID} & \textbf{Demographics}\\
    \midrule
    P1   & 20 years experience, current role as project manager, prior roles as tester, developer, consultant. \\
    P2 & over 20 years experience, current role as requirements and engineer, prior role as developer.\\
    P3   & 15-20 years experience in industry and academia, current role as automation lead,  previously consultant and software and R\&D engineer.\\
    P4  & 9 years experience, current role as manager (development and verification), previously integrator, architect, requirements, safety engineer. \\
    P5  & 41 years of experience, current role as academic, prior role as engineer and consultant. \\
    P6 & 11 years of industrial experience, current role as academic, prior role as designer and requirements engineer.\\
    P7 & 28 years of experience, current role in software engineering, previously across multiple companies.\\
    P8 & around 20 years of experience, current role as senior software engineer, previously at two companies.\\
    P9 & several years of experience, current role as developer and project lead, previously a verification engineer. \\
    P10 & 24 years of experience, current role as manager for a software development team, previously 11 years as a developer, and 9 years as a test engineer. \\
    P11 & around 10 years of experience, current role as a software engineer, previously at three companies. \\
    P12 & 21 years of experience, current role as technology leader, previously roles in several companies and universities. \\
  \bottomrule
    \end{tabular}
\end{table}

Table~\ref{tab:participants_demo} summarizes the 12 interviewees. Experience ranged from 9 to 41 years, with most over 15 years. Six held leadership roles, and others were senior engineers in requirements, development, testing, verification, and configuration management. There are both interviewees with a single-employer tenure and those with multiple-company careers. Three participants described long careers within one firm. At least eight worked across more than two companies, ranging from consulting across many sites, mixed industry-to-academia and consultancy paths, and small to large enterprise mobility. Two interviewees mentioned that during their careers, they interacted with over seven teams. 

\subsubsection{\textbf{Data Analysis}} 

We analyzed the interviews following Braun and Clarke's guidelines for \textit{thematic analysis}~\cite{braun}. 
Initial coding focused on identifying recurring ideas related to the theoretical framework used (perceptions, folklore types, and transmission/impact) while remaining open to unanticipated categories. Coding then proceeded iteratively and collaboratively. After a calibration phase where all authors coded the same interview to establish agreement, 
all other interviews were independently double-coded by rotating pairs of authors. One author then merged the codings for each interview, resolving discrepancies through discussions. 

\begin{table*}[!t]
\scriptsize
\centering
\caption{Selected folklore-related SE literature, sorted by date of publication.}
\label{tab:folklore-summary-extended}
\begin{tabular}{|p{2.2cm}|p{0.9cm}|p{6.6cm}|p{6.6cm}|}
\hline
\textbf{Literature} & \textbf{Mentions Folklore} & \textbf{Contributions} & \textbf{Gaps and Opportunities} \\ \hline
Neumann (1999)~\cite{neumann1999folklore} & Yes & Applies folklore concepts to workplaces; emphasizes rituals and artifacts. & Suggests methods for studying workplace practices; can inform SE studies. \\ \hline
Basili \& Shull (2005)~\cite{basili2005defectfolklore} & Yes & Presents defect folklore as commonly held heuristics; examines their accuracy across studies. & Suggests ways to analyze belief formation and evaluate experience-based knowledge. \\ \hline
Passos et al. (2011)~\cite{passos2022beliefs} & Yes & Uses term ``technical folklore'' to frame belief systems that shape SE practice. & An empirical attempt to connect belief origins, usage, and impact. \\ \hline
Shull (2012)~\cite{shull2012ibelieve} & No & Discusses the role of belief in SE; links beliefs to experience and calls for evidence-based validation. & Highlights the importance of understanding belief persistence; folklore theory could provide additional context. \\ \hline
Spínola et al. (2013)~\cite{spinola2013technicaldebt} & Yes & Defines technical debt folklore as community-based, experience-driven beliefs; surveys practitioners for consensus. & Indicates how shared opinions may influence SE practice; enables examination of belief adoption. \\ \hline
Bossavit (2017)~\cite{bossavit2012leprechauns} & Yes & Deconstructs several well-known SE claims (``leprechauns'') as unverified beliefs passed as facts; supports empirical scrutiny. & Useful to trace belief folklore and myth-making in SE; foundational for defining and demythologizing SE folklore. \\ \hline
Zagalsky (2018)~\cite{zagalsky2018knowledge} & No & Investigates knowledge sharing in developer communities; analyzes media use and human factors. & Offers a foundation for applying folklore concepts to collaborative and knowledge-sharing of narrative aspects of SE. \\ \hline
Méndez Fernández \& Passoth (2019)~\cite{mendez2021interdiscipline} & Yes & Directly connects SE to folklore; describes exemplar folklore in empirical SE; argues folklore is backed up by social mechanisms. & A good starting point in the analysis of folklore grounding; a critique of how conventional wisdom and weak claims persist in SE research and practice.
 \\ \hline
Shrikanth \& Menzies (2020)~\cite{shrikanth2022defectbeliefs} & No & Shows gaps between defect prediction beliefs and experiment results; shows most beliefs are context-dependent. & Opens possibilities for improving belief assessment through monitoring and context-aware adaptation. \\ \hline
Shrikanth et al. (2021)~\cite{shrikanth2021beliefs} & No & Evaluates long-standing SE beliefs; finds most have limited empirical support. & Highlights belief persistence despite lack of evidence; opportunity to interpret this persistence via folklore theory. \\ \hline
Ciancarini et al. (2023)~\cite{ciancarini2023storytelling} & No & Reviews literature on storytelling in SE; identifies functions of stories in collaboration and reasoning. & Basis for interpreting stories as folklore; potential for applying folklore theory to genres and archetypes. \\ \hline
Swillus et al. (2024)~\cite{swillus2024sociotechnical} & No & Explores developers' testing experiences; identifies lived experiences (i.e., rituals and social dimensions of testing). & Entry point to analyze software testing as a folkloric domain with habits, informal norms, and shared identity. \\ \hline
Romano et al. (2024)~\cite{romano2024deadcode} & Yes & A confirmation that removing dead code improves the structure of code as well as resource usage in a specific context. & A single example of how validation can engage with folklore-generation and assumptions in SE.\\ \hline
\end{tabular}%
\end{table*}

We compared codes across interviews to surface common patterns, grouped related codes into candidate themes and subthemes in a digital whiteboard, and refined these through discussions. To aid transparency and traceability, we maintained a shared log of notes, steps, and coding. Coding was both deductive and inductive as well as semantic, allowing single excerpts to map to multiple codes and higher-order themes. We iteratively consolidated overlapping codes and themes. 
We monitored for thematic saturation by tracking the recurrence of themes across interviews and observed that the vast majority are reflected in multiple interviews.

The analysis produced a consolidated set of themes that describe how SE folklore is perceived, manifested, and propagated across organizational settings. The final outputs of the method were anonymized transcripts, a consolidated set of codes, a thematic map with named themes and subthemes, and curated quotations for each theme and subtheme.

\section{Findings}
Our findings are structured as follows: first, we summarize the contributions of selected studies to understanding SE folklore (Section~\ref{sec:lit_review}); second, we present a thematic analysis of curated folklore items (Section~\ref{sec:concepts}); third, we present the thematic analysis of the interviews (Section~\ref{sec:resultsinterview});
finally, we discuss how these findings align with folklore theories and identify conceptual and methodological gaps, proposing a definition of SE folklore (Section~\ref{sec:definition}).  

\subsection{SE Folklore in the Literature}\label{sec:lit_review}

To set a foundation, we reviewed a manually curated sample of research literature that addresses informal knowledge, belief systems, narratives, and ritualized practices within software development. Table~\ref{tab:folklore-summary-extended} presents a selection of thirteen publications that were analyzed according to their explicit discussion of folkloric concepts, their contributions to understanding the folklore of SE, and the specific conceptual or methodological gaps they expose.

While \textit{\ul{the term ``folklore'' is rarely explicitly invoked in the SE literature}}, several studies have addressed the persistence of various claims, myth-like narratives, and socially constructed norms in software development. Some studies explicitly use the term ``folklore'', such as Bossavit~\cite{bossavit2012leprechauns}---who critiques unempirical claims passed down as truths---and Spínola et al.~\cite{spinola2013technicaldebt}, who label diverging practitioner opinions on technical debt as ``technical debt folklore''. Others, such as Neumann~\cite{neumann1999folklore}, provide theoretical grounding from library and information science, applying folklore studies to analyze how rituals and narratives shape professional identity.
Several papers (e.g., Swillus et al.~\cite{swillus2024sociotechnical}, Shrikanth et al.~\cite{shrikanth2021beliefs}, and Ciancarini et al.~\cite{ciancarini2023storytelling}) reveal folklore (e.g., transmission of beliefs and practices or narratives about testing or project management) but do not frame them as folklore or analyze their symbolic and cultural significance. 

In particular, Bossavit~\cite{bossavit2012leprechauns} examined the concept of folklore, arguing that many widely accepted ``truths'' in SE are myths perpetuated through misinterpretation, citation errors, and a lack of empirical verification. They show how anecdotal evidence and subjective opinions have been repeatedly cited as evidence, shaping industry beliefs without robust validation. Bossavit advocates for a more skeptical approach to SE knowledge, emphasizing the need for rigorous scrutiny and historical awareness to prevent the propagation of unfounded claims.

\begin{table*}[!t]
\scriptsize
\centering
\caption{Analysis of identified folklore items. 
Symbolic Meanings: A = Artifact, EV = Espoused Beliefs and Values, BUA = Basic Underlying Assumptions. 
Occupational Groups: DEV = Developers, TST = Testers, MNT = Maintainers, ARC = Architects, PM = Project Managers, ALL = All roles. 
Knowledge Areas: DSN = Design, CNST = Construction, TEST = Testing, MNT = Maintenance, MGMT = Management, PROC = Process, QUAL = Quality, ECON = Economics, PRO = Professional Practice.}
\label{tab:folklore_items_sources}
\begin{tabular}{|l|p{8.5cm}|l|p{1.3cm}|p{1.7cm}|p{2.3cm}|}
\hline
\textbf{ID} &  \textbf{Folklore Item} & \textbf{Narrative Form} & \textbf{Symbolic Meaning} & \textbf{Occupational Groups} & \textbf{Knowledge Areas} \\
\hline
F1 & The vast majority of defects are interface defects~\cite{basili2005defectfolklore}. & Myth & BUA & DEV, TST & CNST, QUAL, DSN \\
\hline
F2 & Object-oriented programming reduces errors and encourages reuse~\cite{shrikanth2021beliefs}. & Myth & EV & DEV & DSN, CNST, QUAL \\
\hline
F3 & Higher software quality results in better productivity~\cite{shrikanth2021beliefs}. & Myth & EV & DEV, PM & PROC, QUAL \\
\hline
F4 & Developer performance varies dramatically (e.g., 10x productivity)~\cite{shrikanth2021beliefs,bossavit2012leprechauns}. & Legend & BUA & DEV, PM & PRO \\
\hline
F5 & Becoming an expert requires at least 5000 hours~\cite{shrikanth2021beliefs}. & Myth & EV & DEV & PRO \\
\hline
F6 & Technical debt is unavoidable in real-world projects~\cite{spinola2013technicaldebt}. & Myth & EV & DEV, ARC & MNT, PROC \\
\hline
F7 & Unintentional technical debt is much more problematic than intentional~\cite{spinola2013technicaldebt}. & Myth & EV & DEV, MNT & MNT \\
\hline
F8 & Technical debt originates from short-term optimization~\cite{spinola2013technicaldebt}. & Myth & EV & DEV, PM & DSN, MNT \\
\hline
F9 & Defect classes follow organization-specific patterns that can be detected within a given context~\cite{basili2005defectfolklore}. & Myth & BUA, EV & TST, DEV & QUAL, PROC \\
\hline
F10 & Developers are more likely to test when they feel ownership over the code~\cite{swillus2024sociotechnical}. & Anecdote & EV & DEV & TEST \\
\hline
F11 & Testing is a tedious burden, far less exciting or rewarding than coding~\cite{swillus2024sociotechnical}. & Anecdote & EV, BUA & DEV & TEST \\
\hline
F12 & If you are not sure what to do, do something and fix it later~\cite{basili2005defectfolklore}. & Myth & BUA & DEV & CNST, MNT \\
\hline
F13 & Personal artifacts and cluttered desks are part of work identity~\cite{neumann1999folklore}. & Artifact & A & DEV & PRO \\
\hline
F14 & Jokes, photocopies, and office memes circulate informally in teams~\cite{neumann1999folklore}. & Humor & A & ALL & PRO \\
\hline
F15 & Dead code removal improves performance and maintainability~\cite{romano2024deadcode}. & Myth & EV & DEV, MNT & CNST, MNT \\
\hline
\end{tabular} 
\end{table*}

This selected body of work highlights a need for a definition of SE folklore that captures the informally transmitted knowledge and practices shared within occupational software development groups. Furthermore, it reveals several gaps and opportunities, including the absence of methods to study folklore elements such as rituals, storytelling, and belief formation in SE contexts. While some studies identify and validate beliefs (e.g., about defects, technical debt, or dead code), others reveal a persistence of unverified or outdated practices. 
Overall, few studies explicitly apply methods or theories from ethnographic studies to the software development domain~\cite{sharp2016role}. Such methods could help explain how cultural narratives, myths, and shared practices influence both researchers and practitioners.
This gap motivates the need to establish SE folklore as a formal concept with descriptive, critical, and cultural explanatory power.

\subsection{Analysis of the Identified Folklore Items}\label{sec:concepts}

Table~\ref{tab:folklore_items_sources} presents an analysis of folklore-related items (F1–F15) derived from the studies in Table~\ref{tab:folklore-summary-extended}. Our intention was not to provide an exhaustive account of folklore items, but to highlight exemplar cases that can be used to examine various dimensions and concepts of SE folklore. The folklore items span a range of topics---beliefs about where defects occur~\cite{basili2005defectfolklore}, the relation between quality, productivity, and experience~\cite{shrikanth2021beliefs}, views on technical debt~\cite{spinola2013technicaldebt}, perceptions such as the ``super-programmer''~\cite{bossavit2012leprechauns}, and the idea that removing dead code improves maintainability and performance~\cite{romano2024deadcode}. 

Most items are framed as myths---generalized beliefs rooted in collective experience---particularly within software construction, maintenance, and quality. These myths typically express espoused values (e.g., F2–F3, F5–F8), but some reflect basic underlying assumptions (e.g., F1, F4, F12), meaning they seem to be taken for granted and are rarely questioned in practice. 

For example, F1 disseminates the idea that the vast majority of defects are interface defects. This item seems to operate as a basic underlying assumption for developers and testers engaged in  construction, design, and quality work. F5 advances the claim that becoming an expert requires at least 5,000 hours. As an espoused value, this item seems to deliver an encouraging message---sustained, focused effort is a route to deep SE competence. Shrikanth et al.'s reassessment indicates that this rule holds only in some cases~\cite{shrikanth2021beliefs}. Novices often match experts in both speed and quality, suggesting that experience adds little unless deliberate practice is in place. 

Some entries are anecdotes (F10–F11)\footnote{In our framing, an anecdote is a short, experience-based story used to illustrate a point. We tagged items as anecdotes when they are situated accounts. For example, F10 and F11 are commentaries made by interview participants in a specific case~\cite{swillus2024sociotechnical}.} or artifacts and humor (F14), yet these still signal cultural identity, emotion, or irony that guide team practices and shape social aspects of their work.  Consider F11, the anecdotal belief that testing is a tedious burden. This item combines both espoused and underlying symbolic layers, indicating perceived tensions around the value of testing (relative to coding) that resonate in the testing area and across team cultures. 

F13---the idea that personal artifacts and cluttered desks are part of identity---is categorized as an item related to ``artifacts'', as this belief reveals how material environments carry symbolic importance in practitioners' self-expression and professional practice. Meanwhile, F4 (the $10x$ developer legend) is built on decades of stories of mythical practitioners. This legend could potentially influence hiring, team composition, and performance expectations, especially among developers and project managers. These examples highlight how one can not only catalog folkloric content but also surface the socio-cultural operations that underpin SE. 


Qualitative analysis of the folklore items reveals several themes. First, \textit{\ul{many beliefs serve as heuristics that simplify reasoning in technically complex situations}}. These include commonly repeated rules of thumb, such as the fact that most defects occur at module interfaces or that unintentional debt is worse than intentional debt, which seem to provide practical guidance, despite such guidance often lacking clear empirical grounding. 

Other folklore items relate to \textit{\ul{identity-related dimensions of SE work}}. The perception that testing is tedious or not rewarding~\cite{swillus2024sociotechnical}, or that a cluttered workspace is a sign of professional identity~\cite{neumann1999folklore}, reflects deeper assumptions about roles, hierarchies, and belonging. Similarly, the narrative of the $10x$ developer could reinforce individualist identity ideals within organizational culture. \textit{\ul{Folklore related to technical debt}} emerges as a particularly rich domain. Cultural beliefs about the inevitability of debt and its connection to short-term decision-making highlight organizational challenges. 

Several folklore items appear to persist despite being challenged. These include beliefs in the universal benefits of specific design patterns and exaggerated claims about variations in developer performance. Their durability may suggest that factors such as cultural alignment, personal experience, intuition, and social reinforcement often outweigh empirical evidence in shaping beliefs. In addition, \textit{\ul{ritual and informality}} also hold significant folkloric meaning. 

\textit{\ul{The ``transmission'' of folklore remains understudied}}. With few exceptions---i.e., via community interactions~\cite{zagalsky2018knowledge} and through shared project and learning practices~\cite{swillus2024sociotechnical}---there is little evidence or direct investigation into how these beliefs are passed from one practitioner to another, whether through mentorship, documentation, onboarding, or socialization. There is a need to trace transmission mechanisms to better understand how these practices persist and evolve across different contexts and through various media.

\subsection{Interview Study Results}\label{sec:resultsinterview}
In this section, we present the interview results organized into practitioners' general perceptions of folklore, concrete folkloric forms (i.e., myths, anecdotes, rituals, artifacts, humor), and their transmission and impact on SE work. The emerging themes contribute directly to what constitutes SE folklore and how it can be defined by grounding the concepts in practitioners' accounts. 

\subsubsection{\textbf{General Perceptions}}
Participants described SE folklore as a lived and practical phenomenon that sits between what practitioners \textit{say} they do and what they \textit{actually} do. Three emerging themes capture these views: \textit{\ul{First Impressions of What Folklore Is}}, \textit{\ul{Practice versus Ideals}}, and \textit{\ul{Cultural Expressions}}.

When asked what folklore meant, participants described it as unwritten, informal, and often second-hand know-how, typically shared through code reviews, onboarding, or hallway talk. Some framed it as myths and jokes, while others saw it as beliefs without direct evidence, overlapping with opinions and rules of thumb.

Because it functions implicitly, folklore is rarely named as folklore; it persists in practice and only becomes visible on reflection or when someone tries to spread it: 
\interviewquote{P7: Stories with monsters in the [forest], but in software engineering, these are beliefs… not supported by data or evidence.} 

Participants perceived much of what circulates as ``{\it common wisdom}'' driven more by opinion than by evidence, yet still shaping choices. They expressed that folklore captures the friction between ideals and practice, with cultures influencing which stories take hold and how they guide action: 

\interviewquote{P12: I think it is that perception and reality and myth and what is actually happening... and also which is the culture, because I believe that [culture] also matters. Even if you want to go against something, the culture forces you to come back and follow the guidelines.} 


\noindent Simultaneously, folklore was valued for its pragmatism, offering workarounds and context that formal processes may lack. 

Folklore was also described as something you can see in how teams communicate. Participants pointed to national style differences in discourse and humor---for example, norms around constant counter-argument or the tolerance for rougher jokes---as shaping which stories get repeated:
\interviewquote{P5: We are a multicultural company, but... overall company culture promotes how knowledge transfer happens... I did feel barriers because of peoples' skill set.} 
\noindent These views suggest that folklore is a constraint and an enabler. 

\subsubsection{\textbf{Myths and Beliefs}}

Our analysis surfaced myths and beliefs. These beliefs act as shortcuts for coordination and justification, but they also disrupt planning, resourcing, and evaluation. We organize findings into seven themes, comprising recurring subthemes.

\smallskip\noindent\textit{\ul{Testing Myths:}} Participants described testing as both marginalized and misunderstood. Testing was frequently framed as ``{\it minority work}'' delegated to the least empowered roles. 
\interviewquote{P1: As a tester, it is easy to be seen as a nitpicker... We will release the product as soon as the testers stop finding problems. So we need to test less... So the testers are the cause of the project delay.} 

At the same time, code coverage is pushed as a KPI, leading to performative activities to ``{\it hit the number}''.  Across teams, testing was positioned as ``{\it less than}'' development---or ``{\it anyone can do it}''---which was perceived as deteriorating test competencies in the organization. One participant mentioned that some practitioners caricatured exploratory testing as quick and freestyle, overlooking the structure and skill it requires. 

Several participants noted automation myths, e.g., test automation being equated with progress. At the same time, automation's limits were understood as: ``{\it automation only covers the easy parts}''. Another participant mentioned the myth ``\textit{AI will replace testers}''. 

\smallskip\noindent\textit{\ul{Process Myths:}} Participants repeatedly encountered ``{\it silver bullet}''-type myths. For example, model-based SE (MBSE) was promoted as a savior, with promises of fit, ease, and immediate benefits. When benefits were delayed, blame shifted to immature teams rather than to poor fit to the context: 

\interviewquote{P7: Model-based development, or MBSE, holds great promise... But it has become a myth that it is suitable for all software.}

Related to SE processes, the myth of a linear, sequential waterfall/V-model process, where activities are completed and never revisited, persisted even in organizations that practiced iterative work. 



\smallskip\noindent\textit{\ul{Hype and Status-Based Adoption:}} Some participants mentioned that some choices were guided by hype and status in the community rather than fit. One participant described half-baked research repackaged as a product and sold via consultancy narratives.  


Participants also discussed the myth that ``{\it newer means better}'', frequently coupled with the idea that all organizations should imitate the practices of the largest tech companies, regardless of fit:  

\interviewquote{P7: ...We must use a new way; this means it is better. Folklore here is that everything new in SE is good. It is natural to have a sense that what is novel is good.} 
\interviewquote{P8: ...this is how [big tech companies] do this thing. So that is why we should do it... do we really have their scaling problems? It does not matter, you know?... It says [big tech] on the box of this technology, so it must be `super awesome'... so you [believe] the myth that it is good because it is from them.} 

\smallskip\noindent\textit{\ul{Developer Myths:}}
Participants reported beliefs related to the practice of development. For example, ``{\it Code is your documentation}'' rationalizes the use of code instead of other software artifacts (e.g., requirements, architectural designs) to document design decisions: 

\interviewquote{P3: So they had very sporadic, unfinished documentation, very, very cryptic... I heard the most famous quote from the previous team, which was like: Code is your documentation... if you want to know how something works, look into the code...} 

Other myths relate to the idea that code---once working---should not be touched again, whether in the case of ``{\it mysteriously working}'' code or code written by ``\textit{perfect senior developers}'': 

\interviewquote{P4: ...we had a system where the documentation included who wrote a certain code, and you could see the name. If the name was someone who had been there for fifteen years, a very senior developer, there was a myth that you do not touch it. It is basically perfect.}


\smallskip\noindent \textit{\ul{Myths Shaping Planning and Resourcing:}} We identified several planning and resourcing narratives that combined classic myths with organization-specific heuristics. The Mythical Man-Month myth was mentioned, where managers added people thinking it would make the project ``{\it go faster}''. Teams described ``{\it multiply by pi}'' estimation as a folk correction for optimism leading to poor estimation: 

\interviewquote{P1: Ask the developer ...How long? ...A week... Project manager says, Multiply by pi... I have seen people do that.} 

\textit{Testers causing delays} was a recurring accusation near release gates, reinforcing some of the testing myths mentioned: 

\interviewquote{P1: We will release the product as soon as the testers stop finding problems. So we need to test less… So the testers are the cause of the project delay. [laugh] This is such an anti-pattern: the deadline is the 1st of December, development is delayed, so you squeeze the testing.} 

The belief that ``{\it there is always a new release}'' was mentioned by a participant, similarly ``{\it it can be fixed in software}'' justified shortcuts and late requirement changes. Other myths also shaped the perception of participants: the ``{\it 10-year rule}''---the idea that a technology must be proven for a decade to be adopted---and the ``{\it $10x$ programmer}'' belief were mentioned as influencing hiring and allocation. Several participants mentioned the belief that their project will be ``{\it bug-free if the process is followed}'' to the letter:

\interviewquote{P7: Tollgate meetings: after these, the software magically works... things fall into place once these tollgate meetings occur. Seems useful sometimes. Creates a false sense of security in project planning...} 

Teams in different organizations seem to oscillate between method-freedom and heavy standards. In addition, the ``\textit{1-week handover myth}'' was noted, suggesting that experience and knowledge can be fully documented and transferred on a schedule. 

\smallskip\noindent \textit{\ul{Agile Myths.}} Agile was reduced to rituals in some cases: ``{\it agile means no requirements and documentation}''. Daily stand-ups should be productive, but participants mentioned that they can turn into coordination and performative meetings. Role misunderstandings seem to persist: ``{\it Scrum eliminates project managers}''. Also, agile adoption beliefs in software-intensive organizations often ignore contextual constraints. 

\smallskip\noindent \textit{\ul{AI in SE Myths.}}
AI was framed as an inevitability. The fear of missing out in AI was perceived as accelerating adoption without addressing problems: ``{\it AI will solve everything}'' supported one-size-fits-all automation. At the same time, ``{\it AI will replace experts}'' fueled narratives around adoption. 

\subsubsection{\textbf{Anecdotes and Legends}}
Our thematic analysis identified four themes related to legends and anecdotal lore.

\smallskip\noindent\textit{\ul{Control and Compliance Lore:}}
Participants described a body of stories about compliance work, where legal and escrow tales circulate as warnings. For example, a customer asked for the escrow password, which exposed a leak and became a cautionary tale. Other stories of resistance to scaled processes, such as ad-hoc releases, were mentioned. 

\smallskip\noindent \textit{\ul{Memory and Cautionary Lore:}}
Practitioners maintain a living memory. For example, some expressed nostalgia for ``{\it the onsite good old days}'' where software was made by small teams in a single location: 

\interviewquote{P3: Ah, the good old times, we would just go there, we would sit, and we would build the system and it would work because it was being made on the spot. But of course this does not scale.} 

Counteracting this, others discussed hype-cycle legends recounting technologies embraced and later rolled back. These stories are used to caution against trend-based adoption. In both cases, memory can act as a decision heuristic.

\smallskip\noindent \textit{\ul{Legendary Bugs and Narrative Bias:}}
Across interviews, several failure stories were repeatedly mentioned, including the ``{\it Denver baggage failure}'', ``{\it Therac-25}'', and other classic failures:

\interviewquote{P5: There was an article in 1994 in Scientific American where they had this practical example of the luggage system at Denver Airport. It delayed the opening of the airport by six months or so and cost zillions of dollars...  What they did not tell was that just a few years after, everything was up and running, and the luggage system paid back the investment and generated a profit.}

Participants acknowledged the spread of narrative bias---failures are retold and amplified, whereas later successes receive little attention. Nevertheless, these legendary bugs are treated as domain-specific lessons that could influence practices. 

\smallskip\noindent \textit{\ul{Cautionary Tales about Testing and Design Debt:}}
A fourth theme centers around how design and testing choices are transformed into folklore. Tales of designs that resist external demands reveal workarounds that satisfy short-term pressure but create hidden histories in artifacts---design decisions that future teams must decode. Hardware and software boundary tales persist as well---decoy circuits are recounted as clever but debt-creating design decisions. In contrast, participants circulated positive narratives about how architecture and design patterns enabled fast change. 

\begin{figure*}[ht]
\centering
\includegraphics[width=0.27\textwidth]{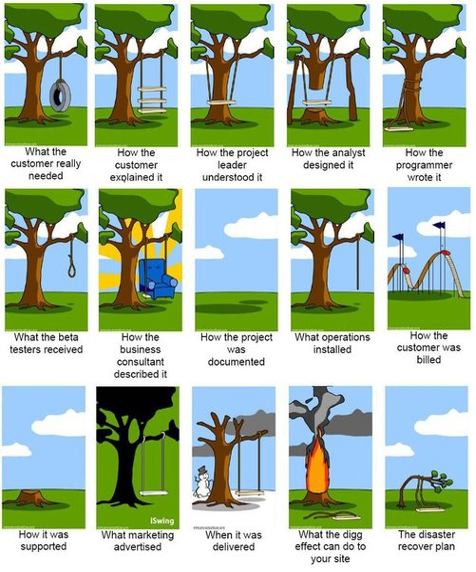}
\includegraphics[width=0.32\textwidth]{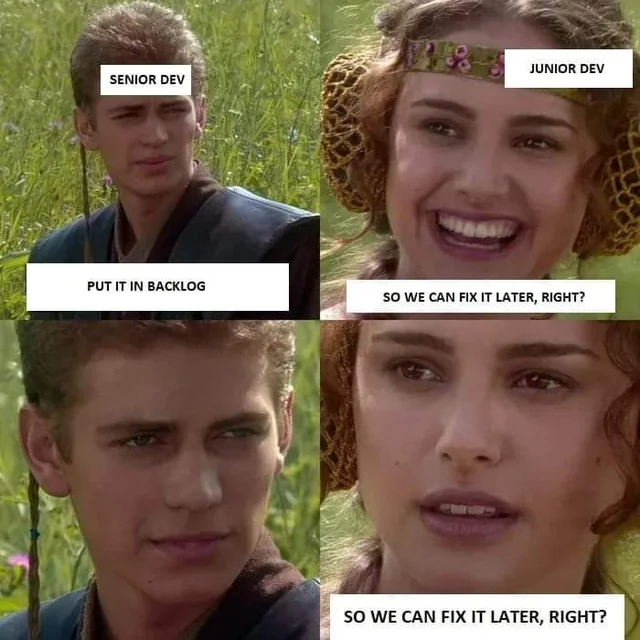}
\includegraphics[width=0.32\textwidth]{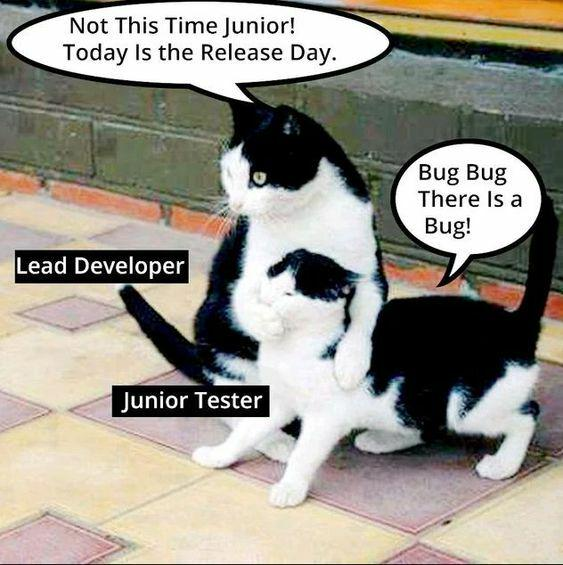}
\Description{Examples of memes mentioned in the interviews. These are explained in the text.}
\caption{Illustrative examples of transmitted memes referenced in the interviews.}
\label{fig:memes} 
\end{figure*}

Testing-related tales list routine development frictions---e.g., testing is never-ending, clashes appear when test code owned by territorial developers (\textit{``my baby''} code) or  \textit{``sacred''} legacy modules are owned by experts. 

\subsubsection{\textbf{Rituals and Practices}}
Across interviews, participants described rituals and practices that pass on know-how, coordinate work, and sometimes shape behavior, clustering into five themes.

\smallskip\noindent \textit{\ul{Reuse Practices:}}
Participants described blind reuse without validation (``{\it it worked there, reuse it here}'') that later required costly rework, and repeated naming conventions that embed hidden histories in artifacts (e.g., cryptic column names or inherited signal labels). Some ``{\it superstitions}'' also persist, such as adding timing delays in test cases or reset habits, which originated in manual testing but are now used in automated tests. 

\smallskip\noindent \textit{\ul{Mentorship Patterns:}} Informal mentorship often takes the form of withholding direct answers to develop further skill:
\interviewquote{P3: ...you could get [ten] questions that are leading you to get the answer... But it was also a very unorthodox way of teaching someone. [...] But yeah, it was a drill to prepare you for the environment when you go to the site... a very unorthodox kind of mentor[ship].}

Also, weekly senior syncs offer a forum where experts share stories, heuristics, and surface patterns. Together, these forms create a channel to transmit knowledge. 

\smallskip\noindent \textit{\ul{Rituals as Social Glue and Morale Building:}}
Teams use formal meetings as bonding opportunities. Stand-ups are used for social sessions; sprints are named after desserts with small rewards and recurring syncs as team-building. These rituals seem to be used to create belonging:

\interviewquote{P3: ...for the sprint name, we would choose some kind of a name for a cookie or some dessert... and the reason was that if we managed to finish above [certain threshold] of the tasks in that sprint, then we would get it at the sprint retrospective or closing meeting.}

\noindent \textit{\ul{Community of Practice:}}
Outside formal roles, engineers sustain their community through voluntary coding meetups and discussions or education around quality: 

\interviewquote{P8: We have a few colleagues that meet up and do some programming online once a week... it is also not typically [a] work-related thing. For instance, let us try out some property-based testing, or something like that. A bit of fun, but also a bit of learning and doing it together.}

These sessions provide places to explore new techniques. In addition, keeping ``{\it quality}'' as a standing topic (e.g., code review checklists, defect narratives) can be used to re-frame perceptions around joint responsibility. 


\smallskip\noindent \textit{\ul{Rituals for Control, Learning, and (Sometimes) Surveillance:}}
Certain practices operate as a form of lightweight control over practitioners. Reviews and tollgates are treated as guarantees of quality and progress. Coffee or lunch breaks function as informal problem-solving sessions that seem to reduce silos and shared spaces were repeatedly cited as places where stories travel. Work-from-home mistrust was perceived as treating visibility as productivity, pushing in-person attendance rituals. Within SE practice, code review is narrated as a form of collective learning; checklists sometimes ensure rigor, but at times are used ``{\it for visibility}''. Participants also mentioned stand-ups as a form of surveillance (``{\it tell me you are working}'') alongside standard planning practices that improved alignment and shortened feedback loops. 

\subsubsection{\textbf{Artifacts and Humor}}

We identified two emerging themes. 

\smallskip\noindent \textit{\ul{Humor as Coping and Critique:}} Practitioners signal difficulty, emotion, and critique through jokes---e.g., the \textit{``backlog equals graveyard''} or \textit{``feature future''} metaphor shows skepticism about using prioritization mechanisms to delay or silently cancel a feature, \textit{``drinking because testing''} frames testing as an ordeal, and ``\textit{trolls explain flaky systems}'' offers a playful story that normalizes intermittent failures. Circulating texts, such as a printed quote sheet (i.e., catchphrases), provide a channel for critique and mentoring. Humor is also a form of resistance---e.g., using a joke to challenge unrealistic safety requirements or self-proclamation of oneself as a ``{\it pain in the ass}'' to show pride in verification and quality gatekeeping efforts. Newcomer socialization is evident in first-day jokes that motivate and tease, introducing newcomers to local expectations. Finally, cross-team banter that can go into blame cultures can be seen as a double-edged sword: it can coordinate developers and testers, for example, but when tensions rise, it can drift into stigma. 

\smallskip\noindent \textit{\ul{Meme Culture}:} Memes, generally in the form of images, are a specific form of humor often shared within teams. Memes resurface when certain events recur, allowing practitioners to learn from and remember past incidents. Memes are often created locally by practitioners for their specific projects or histories, and are often used to relieve hierarchical tensions (e.g., jokes about managers). 

Several general memes were mentioned (some shown in Figure~\ref{fig:memes})---e.g., the ``\textit{swing}'' meme (showing miscommunication about requirements), the ``\textit{silenced junior}'' (two cats fighting, with the junior being silenced to avoid delaying a release), and \textit{``backlog and fix it later''}. One participant also mentioned the ``{\it this is fine}'' (shared experience of living with chaos) and ``\textit{real men test in production}'' (warning about the implications of minimizing testing) memes. 


\subsubsection{\textbf{Transmission and Impact}}
Participants described SE folklore as both shaping choices and structuring how newcomers learn. 

\smallskip\noindent \textit{\ul{Perceived Impact of Folklore:}} Folklore affected decisions and the everyday environment. For example, model-based SE drove adoption decisions and participants emphasized that the effects were both harmful and helpful, depending on the context: 

\interviewquote{P2: MBSE will be a big game changer...[a] decision to introduce MBSE... created a lot of challenges... misunderstanding of when you get  benefits.} 

In routine interactions, folklore served as a way of teaching or venting and was seen as generally helpful, yet this coexisted with an explicit, mostly harmful verdict on folklore due to perceived costs and for instilling non-evidenced beliefs. These doubts can affect which folklore items are introduced to newcomers and sustained.

\smallskip\noindent \textit{\ul{On-boarding Transmission:}} The first transmission seems to occur during on-boarding. Memes used in on-boarding create shared references, often flowing from senior to junior alongside cautionary tales. In some teams, deliberately minimal or cryptic documentation also seems to push newcomers to lean on folklore. This early shaping seems to connect later to patterns.

\smallskip\noindent \textit{\ul{Implicit and Informal Transmission.}} Once inside the team, quiet norms can repress questioning, while humor can influence the team climate. At the same time, small talk can help in building a sense of belonging. Folklore seems to be recognized upon reflection: 

\interviewquote{P4: ...this is something we have been unconsciously seeing and practicing and doing, but we have never formally looked at it or thought about it.}

On-boarding stories continue to circulate, and some beliefs are transmitted in one-to-one discussions. 

\smallskip\noindent \textit{\ul{Transmission Mechanisms:}}
These flows seem to be amplified by where and how practitioners meet.  Rumors become known, especially as they are retold in common areas---e.g., coffee or lunch rooms---which become sites for sharing stories and lessons: 

\interviewquote{P9: People already assume that they know the whole picture... and they start spreading that as if it is the gospel.}

Here, informal leaders and lived experience seem to set norms. However, language and cultural barriers can impede participation, shaping who speaks, who listens, and whose lessons are told.

\smallskip\noindent \textit{\ul{Mediators:}} Verification leads are seen as practitioners who counter misconceptions and can change narratives, particularly where some organizations use experimentation. In other cases, some companies seem to allow beliefs to persist. 

\smallskip\noindent \textit{\ul{Scope and Impact:}} Just a few participants reported little to no SE-specific folklore within their organizations, expressing that many of these stories or practices are general to all workplaces. Another participant mentioned that ``{\it negative folklore}'' seems to fuel public mistrust of software and SE. 

\smallskip\noindent \textit{\ul{Organizational Culture:}} Folklore seems to vary by role and organization, and is influenced by company climate. One participant mentioned that team shuffling is a way of refreshing folklore by moving narratives across groups. The size of the organizations could matter too, with one participant mentioning that small firms they were part of are less ritualistic, while large firms accumulate such rituals more often. 

\smallskip\noindent \textit{\ul{Educational Channels:}}
Computer science courses often leave students thinking ``{\it code solves everything}'' and that attitude seems to carry into practice. Some practitioners, especially juniors, also fixate on specific programming languages, a preference that tends to soften with experience.

\smallskip\noindent \textit{\ul{Community Transmission:}}
Practitioner schools-of-thought and skepticism of academic research can shape what folklore teams accept. External channels (i.e., talks by well-known figures, conferences) propagate beliefs, seeding narratives that reappear in practice.

\smallskip\noindent \textit{\ul{Folklore Transfer:}} Participants mentioned evolving forms of transfer across careers. Early-career ``{\it bucket filling}'' involves practitioners accumulating narratives, they consolidate them mid-career, and veterans become set in their way of working (``{\it we tried that}''):

\interviewquote{P11: They tried something, I do not know, 25 years ago, and then they keep saying `we tried that once. It is not gonna work'... but because of this old story, they do not want to even try...}

\subsection{A Software Engineering Folklore Definition}\label{sec:definition}

Software development in industrial settings, particularly within occupational groups, offers a ground for understanding how folklore manifests in SE environments~\cite{shaw2002prospects}. By applying Alan Dundes' characterization---focused on folk groups, informal transmission, and tradition---and Simon J. Bronner's~\cite{bronner2016folklore} practice-centered description---emphasizing praxis, knowledge in action, and phemic processes---we can explore how software development teams cultivate occupational folklore through practices, rituals, and traditions.

Based on Dundes' description, folklore is ``traditional knowledge transmitted informally within folk groups''~\cite{dundes1965computers}. Knowledge is passed through experience, mentorship, or casual interaction rather than formal instruction, while practices and expressions are repeated with variations across teams and projects. The \textit{occupational group} categorizations presented in Table~\ref{tab:folklore_items_sources} provide a basis for understanding how folklore manifests and functions within SE communities. For example, \textit{\ul{software testers, designers, and developers can form distinct folk groups}}. Folklore in software development can help create and reinforce group identity. Informal communication channels can serve as transmission vehicles for occupational traditions, shaping how teams work and interact. 
Our analysis uses a folklore lens (function, transmission, persistence, variation), but shows only a partial view. Without longitudinal data or cross-community circulation evidence, claims about folklore evolution remain interpretive.

Bronner represents folklore as ``traditional knowledge put into, and drawing from, practice''~\cite{bronner2016folklore}. In this explanation, folklore refers to how individuals enact traditions through their daily activities. Consistent with Bronner's practice lens, Enoiu's essay \cite{enoiu2025essay} illustrates how software testing preconceptions and routines circulate as lore, shaping testers’ activities. Thus, actions and expressions can carry suggestive meanings, shaping cultural identity, and folklore emerges through the actual activity of testing software, not just through verbal communication. As shown in Table~\ref{tab:folklore-summary-extended} and Table~\ref{tab:folklore_items_sources}---as well as the themes surfaced in the interviews---folklore is deeply embedded in software development. Practices like testing, programming, reviews, deployments, and retrospectives often take the form of ritualized actions that carry symbolic meaning and help teams navigate complex work environments. Folklore lies in the praxis of these activities and how they are performed, adapted, and symbolized. In addition, following Bronner's notion of folklore as a communicative process integrated in social practices, our definition should emphasize not just what folklore items say, but how and when they are performed.

We synthesize concepts across (1) the folkloristic characterizations and dimensions, (2) the folklore items from literature, and (3) the thematic results from interviews. Building on the folkloristic, we consider the informality of transmission and praxis as foundational principles, both traditional and emergent. The literature review pinpoints the forms these take in SE as narratives and heuristics. Finally, the interviews show these elements in reality---myths, stories, memes, rituals, and practices that actively shape identity, values, and shared know-how across occupational folk groups and knowledge areas. We propose the following definition:
\vspace{0.2cm}
\begin{mdframed}[linewidth=1pt, linecolor=black, backgroundcolor=gray!10, roundcorner=5pt]
\textbf{Software engineering folklore, comprising informally transmitted, traditional, and emergent narratives, heuristics, and artifacts enacted by practitioners, circulates within occupational folk groups (e.g., developers, testers, and managers) and shapes identity, values, and collective knowledge throughout the socio-technical ecosystem of software development.}
\end{mdframed}
\vspace{0.2cm}
This definition highlights the informal nature of knowledge and practices within occupational SE groups. Future research on this topic should further develop this definition and its underlying dimensions and forms, and validate the folklore concepts.


\section{Implications and Applications of SE Folklore}
To operationalize our definition of SE folklore, one would need to treat a folk item as a unit of analysis and evaluate it against the criteria and analytical dimensions (i.e., form, meaning, transmission, spread, practice, and relevance) to enable comparison and analysis. 

For engineers, managers, and team leads, understanding folklore can support more thoughtful cultural and organizational interventions. Practically, this involves identifying persistent narratives and reflecting on their origins, usefulness, and limitations, recognizing when narratives indicate resistance to change or when they are needed to make implicit cultural norms visible. By viewing folklore not as problematic or negative but as cultural knowledge, teams can examine unhelpful stories or beliefs while preserving practices that are effective in their context and that reflect their values.

For researchers in empirical SE and socio-technical systems, folklore offers a lens for examining the lived realities of software teams. The identification of folklore in SE opens the door to empirical and interpretive studies that investigate the human, cultural, and narrative dimensions of tech work. 
Longitudinal, immersive fieldwork in software teams can surface informal beliefs, rituals, and artifacts that shape practices. 
Such ethnographic studies~\cite{zhang2019ethnographic} are well-suited for uncovering knowledge and group identity.
Retrospective accounts and oral history can show how  SE teams’ stories become folklore and how beliefs shift over time~\cite{bornat2004oral}.

\section{Threats to Validity}

We adapted folkloristics to SE, which may shape how folklore is identified and described; future work should refine these constructs with more practitioner input and comparative analysis among researchers. Our literature review also relied on a small, purposively selected sample and an informal search. 
Therefore, a more systematic protocol that includes grey literature is needed. Further, the interview study and the thematic analyses of the identified folklore items are subject to the authors' biases. To mitigate this threat, the coding was performed independently by all authors, then iteratively validated. Disagreements were discussed, resulting in only a few revisions. In addition, our convenience sampling and limited demographic diversity, particularly a skew toward more experienced participants, may bias the themes toward patterns typical of that group. This limitation may restrict generalizability across other experience levels, roles, and organizational contexts. These limitations define the scope of the current work but do not undermine its core contributions in exploring and defining SE folklore. Addressing them in future research will help build a more comprehensive and actionable understanding of SE folklore.

\section{Conclusions}

Drawing on foundational ideas from folklore studies, literature, and an interview study, we have proposed a working definition that characterizes SE folklore as informally shared, traditional, and evolving narratives integral to everyday SE practice.
As in other fields, folklore can function not only as a source of informal knowledge but also as a means of creating identity group cohesion and managing uncertainty in complex social environments. 
Recognizing the role of SE folklore has both practical and theoretical value. For practitioners, it creates opportunities to reflect on norms, challenge myths, and maintain positive customs. For researchers, it opens up new directions for investigation, including ethnographic and folklore studies, comparative analysis across domains, and methodological tools. 

\begin{acks}
Support was provided by the Software Center project 68 (TRACE) and MONA LISA (ITEA) project funded by Vinnova. Enoiu was also supported through the AI and Society Fellowship (AI@MDU). We thank Alex Cusmaru for his input and valuable discussions on the SE folklore concept and an earlier version of the manuscript.
\end{acks}



\bibliographystyle{ACM-Reference-Format}
\bibliography{vr_ram_2014_pertes}


\end{document}